\documentclass[12pt,a4paper,english]{article}
\usepackage[T1]{fontenc}
\usepackage[latin1]{inputenc}
\usepackage{graphicx}

\makeatletter

\newcommand{\lyxaddress}[1]{
\par {\raggedright #1
\vspace{1.4em}
\noindent\par}
}

\usepackage{a4wide}
\makeatletter 

 \@addtoreset{equation}{section}

 \@addtoreset{figure}{section}

 \@addtoreset{table}{section}

\makeatother

\usepackage{babel}
\makeatother
\begin{document}

\title{Casimir effect in the boundary state formalism}

\author{Z. Bajnok$^{1}$, L. Palla$^{2}$ and \underbar{G. Tak\'acs}$^{1}$}

\date{October 19, 2007}

\maketitle

\lyxaddress{$^{1}$\emph{\small Theoretical Physics Research Group, Hungarian
Academy of Sciences, 1117 Budapest, P\'azm\'any P\'eter s\'et\'any 1/A, Hungary}
\\
$^{2}$\emph{\small Institute for Theoretical Physics, E\"otv\"os University,
1117 Budapest, P\'azm\'any P\'eter s\'et\'any 1/A, Hungary}}

\begin{abstract}
Casimir effect in the planar setting is described using the boundary
state formalism, for general partially reflecting boundaries. It is
expressed in terms of the low-energy degrees of freedom, which provides
a large distance expansion valid for general interacting field theories
provided there is a non-vanishing mass gap. The expansion is written
in terms of the scattering amplitudes, and needs no ultraviolet renormalization.
We also discuss the case when the quantum field has a nontrivial vacuum
configuration. 
\end{abstract}

\section{Introduction}

The Casimir effect can be considered as a response of the ground state
in a quantum field theory to the presence of boundary conditions.
Therefore it is natural to seek a relation to the approach known as
boundary quantum field theory started in two-dimensional space-time
by the seminal paper of Ghoshal and Zamolodchikov \cite{GZ}. Recently
we have developed and extended this formalism to quantum field theories
in arbitrary space-time dimensions and applied it to the Casimir effect 
\cite{bLSZ,BQFT,ourcas,bstate}. Here we give a short review of our results.

\section{Boundary state formalism}

\subsection{The concept of the boundary state}

Following \cite{bstate} we consider an Euclidean quantum field theory
of a scalar field $\Phi$ defined in a $D+1$ dimensional half space-time,
parameterized as $(x\leq0,y,\vec{r})$, in the presence of a codimension
one flat boundary located at $x=0.$ The correlation functions defined
as \[
\langle\Phi(x_{1},y_{1},\vec{r}_{1})\dots\Phi(x_{N},y_{N},\vec{r}_{N})\rangle=\frac{\int\,\mathcal{D}\Phi\:\Phi(x_{1},y_{1},\vec{r}_{1})\dots\Phi(x_{N},y_{N},\vec{r}_{N})\, e^{-S[\Phi]}}{\int\,\mathcal{D}\Phi\, e^{-S[\Phi]}}\]
contain all information about the theory. The measure in the functional
integral is provided by the classical action\[
S[\Phi]=\int d\vec{r}\int_{-\infty}^{\infty}dy\left[\int_{-\infty}^{0}dx\left(\frac{1}{2}(\vec{\nabla}\Phi)^{2}+U(\Phi)\right)+U_{B}(\Phi(x=0,y,\vec{r}))\right]\]
 which determines also the boundary condition via the boundary potential
$U_{B}$:\[
\left.\partial_{x}\Phi\right|_{x=0}=\left.-\frac{\delta U_{B}(\Phi)}{\delta\Phi}\right|_{x=0}\]
Here we assume for simplicity that the boundary term does not depend
on the temporal (i.e. $y$) derivative of $\Phi$, which means that there are no 
boundary degrees of freedom with a temporal dynamics independent of the bulk 
(it may depend on derivatives with respect to $\vec{r}$, which is the reason 
for the variational derivative $\delta$). The bulk interaction
$U$ is constrained by the requirement that the bulk spectrum must
possess a mass gap $m$. 

This Euclidean quantum field theory can be considered as the imaginary
time version of two different Minkowskian quantum field theories.
We can consider $t=-iy$ as Minkowskian time and so the boundary is
located in space providing nontrivial boundary condition for the field
$\Phi.$ The space of states in this Hamiltonian description is the
boundary Hilbert space $\mathcal{H}_{B}$ determined by the configurations
on the equal time slices. $\mathcal{H}_{B}$ contains multi-particle
states and is built over the vacuum state, obtained in the presence of the boundary 
condition ($\vert0\rangle_{B}$), by the successive application of particle creation operators%
\footnote{One can also introduce particle-like excitations confined to the boundary
\cite{BQFT} ('surface plasmons'), but for simplicity we do not consider
them here.%
}. In the asymptotic past the particles do not interact and behave
as free particles travelling towards the boundary; thus \[
\mathcal{H}_{B}=\left\{ a_{in}^{+}(k_{1},\vec{k}_{1})\dots a_{in}^{+}(k_{N},\vec{k}_{N})\vert0\rangle_{B}\quad,\qquad k_{1}\geq\dots\geq k_{N}>0\right\} \]
where the operator $a_{in}^{+}(k,\vec{k})$ creates an asymptotic
particle of mass $m$ with transverse (i.e. $x$-directional) momentum $k$ and parallel (i.e. parallel to the boundary) momentum
$\vec{k}$. The corresponding energy is $\omega(k,\vec{k})=\sqrt{k^{2}+\vec{k}^{2}+m^{2}}=\sqrt{k^{2}+m_{\mathrm{eff}}(\vec{k})^{2}}$
where $m_{\mathrm{eff}}(\vec{k})=\sqrt{\vec{k}^{2}+m^{2}}$ is the
effective mass of a particle with parallel momentum $\vec{k}$ as
seen in the two-dimensional space-time formed by $t$ and $x$. Instead
of $k$, we shall also frequently use the rapidity parameter $\vartheta$
defined by\begin{equation}
\omega=m_{\mathrm{eff}}(\vec{k})\cosh\vartheta\quad,\quad k=m_{\mathrm{eff}}(\vec{k})\sinh\vartheta\label{eq:ddimrapidity}\end{equation}
 In the Heisenberg picture the time evolution of the field\[
\Phi(x,t,\vec{r})=e^{iH_{B}t}\Phi(x,0,\vec{r})e^{-iH_{B}t}\]
 is generated by the following boundary Hamiltonian \begin{equation}
H_{B}=\int d\vec{r}\left[\int_{-\infty}^{0}dx\left(\frac{1}{2}\Pi_{t}^{2}+\frac{1}{2}(\partial_{x}\Phi)^{2}+\frac{1}{2}(\vec{\partial}\Phi)^{2}+U(\Phi)\right)+U_{B}(\Phi(x=0))\right]\label{eq:boundaryham}\end{equation}
The correlator can then be understood as the matrix element \[
\langle\Phi(x_{1},y_{1},\vec{r}_{1})\dots\Phi(x_{N},y_{N},\vec{r}_{N})\rangle=\,_{B}\langle0\vert T_{t}\left(\Phi(x_{1},t_{1},\vec{r}_{1})\dots\Phi(x_{N},t_{N},\vec{r}_{N})\right)\vert0\rangle_{B}\]
where $T_{t}$ denotes time ordering with respect to time $t$, and
the vacuum $\vert0\rangle_{B}$ is normalized to $1$.

\begin{figure}
\begin{centering}
\includegraphics[scale=0.5]{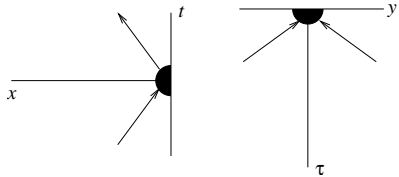}
\par\end{centering}
\caption{\label{fig:The-two-channels} The two Hamiltonian descriptions, with
a representation of the amplitudes $R$ and $K^{2}$.}
\end{figure}

The formulation of asymptotic states and fields, together with the
relevant reduction formulae (which generalize the LSZ approach to
boundary QFT) was given in \cite{bLSZ,BQFT}. In \cite{BQFT} we also
gave the appropriate generalization of Landau equations, Coleman-Norton
interpretation and Cutkosky rules, together with an example of one-loop
renormalization of boundary interaction (where we considered the case
of sine-Gordon model in two space-time dimensions). Elastic reflection
of a particle from the boundary (see figure \ref{fig:The-two-channels})
is a process with one particle of energy $\omega$ and parallel momentum
$\vec{k}$ both in the incoming and outgoing state%
\footnote{Energy and parallel momentum are conserved due to the unbroken translation
invariance in the directions parallel to the boundary.%
}, whose transverse momentum $k$ changes sign. Its amplitude is the
\emph{reflection factor} $R(\omega,\vec{k})$ which can only depend
on $|\vec{k}|$, as a result of rotational invariance in the directions
parallel to the boundary; it is not necessarily unitary due to the
possible existence of inelastic processes.%

Alternatively we can also consider $\tau=-ix$ as Minkowskian time,
as depicted in figure \ref{fig:The-two-channels}. In this case the
boundary is located in time and we can use the usual infinite volume
Hamiltonian description. The Hilbert space is the bulk Hilbert space
$\mathcal{H}$ spanned by multi-particle \emph{in} states \[
\mathcal{H}=\left\{ A_{in}^{+}(\kappa_{1},\vec{k}_{1})\dots A_{in}^{+}(\kappa_{N},\vec{k}_{N})\vert0\rangle\quad,\qquad k_{1}\geq\dots\geq k_{N}\right\} \]
where $\kappa$ is the momentum in the $y$ direction, and the energy
corresponding to the time direction is given by $\omega(\kappa,\vec{k})=\sqrt{m^{2}+\kappa^{2}+\vec{k}^{2}}$.
One can again use a rapidity parametrization in this channel defined
by\begin{equation}
\kappa=m_{\mathrm{eff}}(\vec{k})\sinh\vartheta\quad,\quad\omega=m_{\mathrm{eff}}(\vec{k})\cosh\vartheta\label{eq:closedrapidity}\end{equation}
Time evolution \[
\Phi(\tau,y,\vec{r})=e^{iH\tau}\Phi(0,y,\vec{r})e^{-iH\tau}\]
is generated by the bulk Hamiltonian \begin{equation}
H=\int d\vec{r}\int_{-\infty}^{\infty}dy\left(\frac{1}{2}\Pi_{\tau}^{2}+\frac{1}{2}(\partial_{y}\Phi)^{2}+\frac{1}{2}(\vec{\partial}\Phi)^{2}+U(\Phi)\right)\label{eq:bulkham}\end{equation}
and the boundary appears in time as a final state in calculating the
correlator: \[
\langle\Phi(x_{1},y_{1},\vec{r}_{1})\dots\Phi(x_{N},y_{N},\vec{r}_{N})\rangle=\langle B\vert T_{\tau}\left(\Phi(\tau_{1},y_{1},\vec{r}_{1})\dots\Phi(\tau_{N},y_{N},\vec{r}_{N})\right)\vert0\rangle\]
The state $\langle B\vert$ is called the boundary state, which is
an element of the bulk Hilbert space and is defined by the equality
of the two alternative Hamiltonian descriptions \[
\langle B\vert T_{\tau}\left(\Phi(\tau_{1},y_{1},\vec{r}_{1})\dots\Phi(\tau_{N},y_{N},\vec{r}_{N})\right)\vert0\rangle=\,_{B}\langle0\vert T_{t}\left(\Phi(x_{1},t_{1},\vec{r}_{1})\dots\Phi(x_{N},t_{N},\vec{r}_{N})\right)\vert0\rangle_{B}\]
where the correspondence is valid if $(i\tau,y)$ is identified with
$(x,it)$. Using asymptotic completeness the boundary state can be
expanded in the basis of asymptotic \emph{in} states as%
\footnote{The bars on top of the $K$ coefficients indicate that the above expansion
is that of the conjugate ({}``bra'') boundary state.%
} \begin{eqnarray}
\langle B\vert & = & \langle0\vert\Bigg\{1+\bar{K}^{1}A_{in}(0,0)\label{eq:bstate_expanded}\\
 &  & +\int_{0}^{\infty}\frac{d\kappa}{2\pi}\int\frac{d^{D-1}\vec{k}}{(2\pi)^{D-1}\omega(\kappa,\vec{k})}\bar{K}^{2}(\kappa,\vec{k})A_{in}(-\kappa,-\vec{k})A_{in}(\kappa,\vec{k})+\dots\Bigg\}\nonumber \end{eqnarray}
which we refer to as the cluster expansion for the boundary state
(where due to translational invariance only bulk multi-particle states
with zero total momentum can appear).

\subsection{Relation between the two channels: $K^{1}$ and $K^{2}$ in terms
of $R$}

The one-point function of the field, due to unbroken Poincar\'e symmetry
along the boundary, only has a nontrivial dependence on $x$:\[
\,_{B}\langle0\vert\Phi(x,t,\vec{r})\vert0\rangle_{B}=G_{bdry}^{1}(x)\]
which corresponds to a nontrivial vacuum configuration in the presence of
the boundary condition. The leading
asymptotic behaviour for $x\rightarrow-\infty$ is given by \cite{bstate}
\begin{equation}
\,_{B}\langle0\vert\Phi(x,t,\vec{r})\vert0\rangle_{B}=\langle0\vert\Phi(0)\vert0\rangle+\bar{g}e^{mx}\label{eq:1ptas}\end{equation}
where $\langle0\vert\Phi(0)\vert0\rangle$ is the vacuum expectation
value in the bulk and $\bar{g}$ is a parameter which is characteristic
of the boundary condition (and also of the field $\Phi$). We recall
that $\vert0\rangle_{B}$ is the ground state of the boundary system
which means that there are no bulk excitations present and the boundary
itself is in its ground state. The absence of bulk excitations is
important for the above asymptotics to be valid; however, (\ref{eq:1ptas}) 
also holds when the boundary is excited ('surface plasmons').

Using the
property of the interpolating field $\Phi$\[
\langle0|\Phi(0)|A(\underbar{k}=\underbar{0})\rangle=\sqrt{\frac{Z}{2}}\]
where $Z$ is the bulk wave function renormalization constant ($0\leq Z<1$),
and from the cluster expansion (\ref{eq:bstate_expanded}) one obtains
the relation%
\footnote{Note that this relation remains valid if the Lagrangian field $\Phi$
is replaced by any bulk interpolating field for the asymptotic particles
and its appropriate wave function renormalization $Z$; in that case
$\bar{g}$ also needs to be replaced by another constant which corresponds
to the field considered.%
}\[
\bar{g}=\sqrt{\frac{Z}{2}}\bar{K}^{1}\]
On the other hand, the existence of nontrivial vacuum expectation
value for the field is generally related to a singularity of the reflection
factor at the particular kinematical point $\vec{k}=0$, $\omega=0$
(i.e. $k=im$ or equivalently $\vartheta=i\pi/2$). In our paper \cite{bstate}
it was shown that this singularity takes the following form \begin{equation}
R(\omega,\vec{k})\sim-\frac{mg^{2}/2}{\omega}(2\pi)^{D}\delta(\vec{k})\label{eq:2pt_sing_assumption}\end{equation}
with $g$ parametrizing its strength. Using the cluster property of
local quantum field theory we proved the following relation \[
\bar{g}=\frac{g}{2}\sqrt{\frac{Z}{2}}\]
valid for general quantum field theories, which yields the expression
of $\bar{K}^{1}$ in terms of $g$:\[
\bar{K}^{1}=\frac{g}{2}\]
This extends a relation previously conjectured in the case of two-dimensional
integrable field theories \cite{dptw_onepoint,bluscher}. In the two-dimensional
case, there is no parallel momentum $\vec{k}$ and the rapidity parametrization
(\ref{eq:ddimrapidity}) takes the form\begin{equation}
\omega=m\cosh\vartheta\quad,\quad\kappa=m\sinh\vartheta\label{eq:2dimrapidity}\end{equation}
As a result, the singularity (\ref{eq:2pt_sing_assumption}) corresponds
to a pole \cite{GZ}\[
R(\vartheta)=\frac{ig^{2}/2}{\vartheta-i\pi/2}\]
Let us now turn to $\bar{K}^{2}$. Using the reduction formulae derived
in \cite{bstate} the relation to $R$ can be obtained as follows:
\[
\bar{K}^{2}(\kappa,\vec{k})=R(\omega\to-i\kappa,\vec{k})\]
This can be written using the rapidity parametrizations (\ref{eq:ddimrapidity},\ref{eq:closedrapidity})
as%
\footnote{Note that the rapidity arguments on the two sides of the equality
are conceptually different, since they correspond to the kinematical
variables of two different channels as defined in (\ref{eq:ddimrapidity})
and (\ref{eq:closedrapidity}). We can consider them related by analytic
continuation.%
}\[
\bar{K}^{2}\left(\vartheta,\vec{k}\right)=R\left(i\frac{\pi}{2}+\vartheta,\vec{k}\right)\]
and this relation fits very well with the pictorial representation
in figure \ref{fig:The-two-channels}. In two space-time dimensions
this is the same as the relation obtained by Ghoshal and Zamolodchikov%
\footnote{They also noted that the relation between the two channels can be
considered as a generalization of the well-known crossing symmetry
to quantum field theories with boundary.%
} in \cite{GZ}. We remark that when the theory in the bulk is free
and the reflection is elastic, the boundary state can be written in
a closed form%
\footnote{In 1+1 dimensions this can be extended to any integrable QFT with
integrable boundary condition \cite{GZ}. %
}\begin{eqnarray}
\langle B\vert & = & \langle0\vert\exp\Bigg\{\bar{K}^{1}A_{in}(0,0)\label{eq:B_exp}\\
 &  & +\int_{0}^{\infty}\frac{d\kappa}{2\pi}\int\frac{d^{D-1}\vec{k}}{(2\pi)^{D-1}\omega(\kappa,\vec{k})}\bar{K}^{2}(\kappa,\vec{k})A_{in}(-\kappa,-\vec{k})A_{in}(\kappa,\vec{k})\Bigg\}\nonumber \end{eqnarray}

\section{Defects and defect operators}

\begin{figure}
\begin{centering}
\includegraphics[scale=0.5]{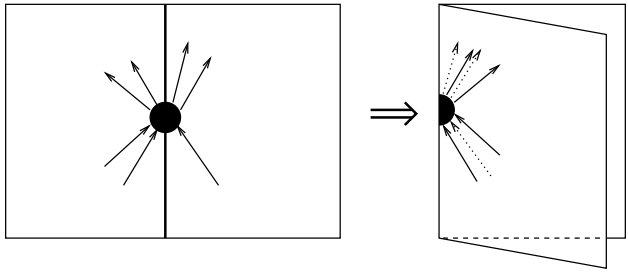}
\par\end{centering}
\caption{\label{fig:The-folding-trick} The folding trick, illustrated for
a generic defect scattering process}
\end{figure}

Boundary conditions considered in the context of the Casimir effect
generally allow transmission as well, and such boundaries are called
'defects'. A suitable generalization of the boundary state formalism
can be obtained by a folding trick depicted in figure \ref{fig:The-folding-trick},
which maps the defect into a boundary system \cite{defect}. Suppose
now that a defect is located at $x_{0}$. In the crossed channel (where
time flows in the $x$ direction) it can be represented by a defect
operator which acts from the bulk Hilbert space of the $x<x_{0}$
system into that of the $x>x_{0}$ system%
\footnote{On the two sides of the defect, the bulk theories may differ; in general,
a defect can be an interface between very different quantum field
theories (as an example one can consider the electromagnetic field
in the presence of an interface between two drastically different physical
media).%
}. %
\begin{figure}
\begin{centering}
\includegraphics[scale=0.7]{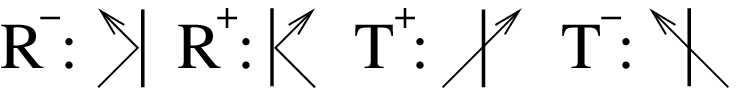}
\par\end{centering}
\caption{\label{fig:One-particle-defect-amplitudes} One-particle defect amplitudes}
\end{figure}
Let us denote the operator creating the particle for the $x<x_{0}$
domain as $A_{1}^{\dagger}$ while for the $x>x_{0}$ domain as $A_{2}^{\dagger}$.
There are now four one-particle reflection amplitudes, shown in figure
\ref{fig:One-particle-defect-amplitudes}. Two of them are denoted
$R^{\pm}$ and preserve the species number $1,2$, corresponding in
the defect picture to reflections on the left and on the right side,
respectively. The other two, $T^{\pm}$ are the ones changing $1$
into $2$ and $2$ into $1$, and in the defect picture they describe
transmission from left to right and right to left, respectively. These
can be conveniently put together into a defect matrix %
\footnote[1]{$D$ is not necessarily unitary, since we allow for inelastic scattering
processes creating and annihilating particles.%
} \[
D(\vartheta,\vec{k})=\left(\begin{array}{cc}
R^{+}(\vartheta,\vec{k}) & T^{-}(\vartheta,\vec{k})\\
T^{+}(\vartheta,\vec{k}) & R^{-}(\vartheta,\vec{k})\end{array}\right)\]
Using the folding map to the boundary system we obtain the defect
operator \cite{defect} as%
\footnote{For the sake of simplicity here we omitted possible one-particle terms
corresponding to nontrivial vacuum configurations, but their inclusion
using the folding trick is straightforward.%
} \begin{eqnarray}
D & = & 1+\int_{\-\infty}^{\infty}\frac{d\vartheta}{4\pi}\int\frac{d^{D-1}\vec{k}}{\left(2\pi\right)^{D-1}}\biggl(R^{+}\Bigl(\frac{i\pi}{2}-\vartheta,\vec{k}\Bigr)A_{1}^{\dagger}(-\vartheta,-\vec{k})A_{1}^{\dagger}(\vartheta,\vec{k})+\label{eq:D_expansion}\\
 &  & \hspace{0cm}T^{+}\Bigl(\frac{i\pi}{2}-\vartheta,\vec{k}\Bigr)A_{1}^{\dagger}(-\vartheta,-\vec{k})A_{2}(-\vartheta,-\vec{k})+T^{-}\Bigl(\frac{i\pi}{2}-\vartheta,\vec{k}\Bigr)A_{1}(\vartheta,\vec{k})A_{2}^{\dagger}(\vartheta,\vec{k})+\nonumber \\
\nonumber \\ &  & \hspace{0cm}R^{-}\Bigl(\frac{i\pi}{2}-\vartheta,\vec{k}\Bigr)A_{2}(\vartheta,\vec{k})A_{2}(-\vartheta,-\vec{k})\biggr)+\mbox{terms with more than two particles}\nonumber \end{eqnarray}
With the same conditions as for the boundary state (trivial bulk scattering,
and elasticity for the combined one-particle reflection/transmission
amplitude) the defect operator can be summed up into an exponential
form similar to (\ref{eq:B_exp}), as discussed in \cite{ourcas}.

\section{Derivation of Casimir energy}

We now turn to the derivation of Casimir energy of a $D+1$ dimensional
scalar field $\Phi(t,x,\vec{y})$ in a domain of width $L$ in $x$
(for details see \cite{ourcas,bstate}). Consider two defects located
at a distance $L$ with defect matrices $D_{1}$ and $D_{2}$. The
ground state eigenvalue of the Hamiltonian $H_{B}$ in (\ref{eq:boundaryham})
can be evaluated via the partition function. Compactifying all infinite
(temporal and spatial) dimensions (i.e. the $D$ extensions perpendicular
to $x$) to circles with perimeter $T$ we can evaluate the partition
function in two different ways \cite{ourcas}:\[
Z(L,T)=\mathrm{Tr}_{\mathcal{H}_{B}}\mathrm{e}^{-TH_{B}}=\left\langle 0\right|D_{1}\mathrm{e}^{-LH}D_{2}\left|0\right\rangle \]
where $H$ is the bulk Hamiltonian (\ref{eq:bulkham}) in the $x$
channel in the domain between the two defects and $\left|0\right\rangle $
is the corresponding bulk vacuum state. Inserting a complete set of
bulk asymptotic states we obtain\[
Z(L,T)=\mathrm{e}^{-LE_{0}}\sum_{n}\left\langle 0\right|D_{1}\left|n\right\rangle \left\langle n\right|D_{2}\left|0\right\rangle \mathrm{e}^{-L(E_{n}-E_{0})}\]
Normalizing the bulk ground state energy $E_{0}$ to $0$, the first
few terms can be written explicitly as\begin{eqnarray*}
1 & + & \sum_{\vartheta,\vec{k}}\sum_{\vartheta',\vec{k}}\left\langle 0\right|D_{1}\vert\vartheta,\vec{k};\vartheta',\vec{q}\rangle\langle\vartheta,\vec{k};\vartheta',\vec{q}\vert D_{2}\left|0\right\rangle \mathrm{e}^{-L\left(m_{\mathrm{eff}}(\vec{k})\cosh\vartheta+m_{\mathrm{eff}}(\vec{q})\cosh\vartheta'\right)}\\
 & + & O(\mathrm{e}^{-3mL})\end{eqnarray*}
The term $1$ is the contribution from the vacuum ($\left|n\right\rangle =\left|0\right\rangle $),
the next term comes from two-particle terms in (\ref{eq:D_expansion})
and the higher-order corrections come from the higher multi-particle
terms. This is a sort of cluster expansion similar to the one used
in \cite{bluscher}, valid for large values of the volume $L$. Finite
volume restricts the momenta to $\kappa=\frac{2\pi}{T}n$ and $k_{i}=\frac{2\pi}{T}n_{i}$,
and the normalization of the creation operators becomes\[
[A_{in}(\kappa,\vec{k}),A_{in}^{+}(\kappa^{'},\vec{k}^{'})]=T^{D}\omega(\kappa,\vec{k})\delta_{\kappa,\kappa^{'}}\delta_{\vec{k},\vec{k}^{'}}\]
The ground state (Casimir) energy (per unit transverse area) can be
extracted from the partition function as \[
E(L)=-\lim_{T\rightarrow\infty}\frac{1}{T^{D}}\log Z(L,T)\]
The result is\begin{eqnarray}
E(L) & = & -\int_{-\infty}^{\infty}\frac{d\vartheta}{4\pi}\,\cosh\vartheta\int\frac{d^{D-1}\vec{k}}{(2\pi)^{D-1}}m_{\mathrm{eff}}(\vec{k})\times\label{eq:leading_order}\\
 &  & R_{1}^{-}\Bigl(\frac{i\pi}{2}+\vartheta,\vec{k}\Bigr)R_{2}^{+}\Bigl(\frac{i\pi}{2}-\vartheta,\vec{k}\Bigr)\mathrm{e}^{-2m_{\mathrm{eff}}(\vec{k})L\cosh\vartheta}+\dots\nonumber \end{eqnarray}
The correction terms correspond to higher particle terms in the expansion
(\ref{eq:D_expansion}) of the defect operator $D$ and include the
amplitudes of reflection/transmission processes involving more than
one particle in at least one of the asymptotic states. These can be
computed (together with the reflection factors $R^{\pm}$) e.g. using
a BQFT formulation as the one presented in \cite{BQFT}, but it is
obvious that they are suppressed by a factor $\mathcal{\mathrm{e}}^{-mL}$
with respect to the leading order term due to the presence of at least
one additional particle in the corresponding term of the expansion
of the defect operator $D$. Note that (\ref{eq:leading_order}) is
applicable in the presence of nontrivial bulk and boundary interactions:
their effects at leading order are contained in the reflection factors
$R^{\pm}$, so as long as there is some theoretical or experimental
input from which these can be determined the leading order contribution
can be evaluated.

In the elastic case the expansion can be summed
up \cite{ourcas}:%
\footnote{We remark that the usual zero mode summation method leads to the same
result, as indicated in Appendix A of \cite{ourcas}. %
}\begin{eqnarray}
E(L) & = & \int_{-\infty}^{\infty}\frac{d\vartheta}{4\pi}\,\cosh\vartheta\int\frac{d^{D-1}\vec{k}}{(2\pi)^{D-1}}m_{\mathrm{eff}}(\vec{k})\times\label{eq:alt3}\\
 &  & \log\biggl(1-R_{1}^{-}\Bigl(\frac{i\pi}{2}+\vartheta,\vec{k}\Bigr)R_{2}^{+}\Bigl(\frac{i\pi}{2}-\vartheta,\vec{k}\Bigr)\mathrm{e}^{-2m_{\mathrm{eff}}(\vec{k})L\cosh\vartheta}\biggr)\nonumber \end{eqnarray}
Let us now calculate the ground state energy in the presence of nontrivial
vacuum configuration of the field. For simplicity we suppose that
the boundary is totally reflective. Compactifying the other directions
again to circles of perimeter $T$ with periodic boundary conditions we obtain
\[
Z(L,T)=\langle B_{\alpha}\vert\mathrm{e}^{-LH}\vert B_{\beta}\rangle=\sum_{n}\frac{\langle B_{\alpha}\vert n\rangle\langle n\vert B_{\beta}\rangle}{\langle n\vert n\rangle}\mathrm{e}^{-E_{n}L}\]
The leading finite size correction to the ground state energy for
large $L$ is now given by one-particle terms, and the ground
state energy per transverse area (at leading order in $L$) has the
form \cite{bstate}
\begin{equation}
E_{0}^{\alpha\beta}(L)=-m\bar{K}_{\alpha}^{1}K_{\beta}^{1}\mathrm{e}^{-mL}+\dots\label{Cas1}\end{equation}
For partially reflecting boundaries (i.e. defects) the appropriate
$K^{1}$ coefficient is the one-particle coupling of the defects evaluated
in the domain between them. If one of the $K^{1}$-s is zero then
the leading correction comes from two-particle states, and is identical
to (\ref{eq:leading_order}).

\section{Summary and discussion}

A very appealing property of the boundary state approach is the universality
of the formulae (\ref{eq:leading_order}) and (\ref{eq:alt3}). In
\cite{ourcas} we showed that the latter indeed reproduces all the
results previously known for the planar situation, including the famous
Lifshitz formula \cite{lifshitz} (it also provides a way to compute
new cases easily, as we demonstrated for a massive scalar field with
Robin boundary condition). 

Another important point is that this approach formulates the Casimir
effect from an infrared viewpoint. Standard derivations of the Casimir
effect solve the microscopic field theory. This necessitates tackling
diverse issues such as renormalization, and also the possibility that
the infrared (long distance behaviour) may be quite different from
the microscopic description of the theory (as is the case for example
in QCD). Formula (\ref{eq:leading_order}) expresses the effect in
terms of the asymptotic particles%
\footnote[1]{Indeed it can be thought of as an expansion in the number of 
virtual particles exchanged between the defects. %
}, and provides a long distance expansion for Casimir energy. 

Our results are consistent with the philosophy behind the more recent
approach by Emig et. al. \cite{jaffe}, the origins of which can be found 
in the earlier papers \cite{balian,feinberg,kenneth}. 
From this viewpoint the Casimir effect
is an interaction of fluctuating surface charge densities, and therefore
it does not logically imply the existence of (astronomically large)
zero point energies because the bulk energy density can be trivially
discarded. In the boundary state approach the surface is characterized
by the coefficients in the cluster expansion of the boundary state
(\ref{eq:bstate_expanded}) (or, more generally, the defect operator
(\ref{eq:D_expansion})). Both approaches give manifestly finite results,
with no ultraviolet divergences whatsoever. There are some differences, 
however. While the boundary state approach only works easily for the planar
case, their methods can be used for general geometries. On the other
hand, the approach of \cite{jaffe} is only formulated for free field theories 
with linear boundary conditions since it relies heavily on the computation of Gaussian
path integrals, while in the boundary state approach the expansion can be written
down for interacting field theories with nonlinear boundary conditions,
in terms of their long distance scattering data. The fact that the
path integral is Gaussian also gives Emig et al. the ability to tackle
theories with zero mass gap, which is only possible in the boundary
state approach whenever the expansion can be resummed into the form
(\ref{eq:alt3}). The boundary state approach, on the other hand,
provides access to highly nontrivially interacting theories with a
mass gap (a prominent example of which is QCD), provided the relevant
scattering data are determined e.g. from lattice field theory (we
remark that it is also highly successful in two-dimensional integrable
field theories where exact scattering amplitudes are known).

It is important to note that the restriction of the boundary state
approach to the planar case comes from the fact that the high symmetry
of the planar situation is exploited to relate the boundary states
(or defect operators) to the scattering data, therefore it is not
a restriction inherent in any theoretical principle. Finally we remark
that the results (\ref{eq:leading_order},\ref{eq:alt3}) automatically
include the contribution of states localized to the defects ('surface
plasmons') as discussed in \cite{ourcas}.

\end{document}